\newcommand{\f}[1]{Fig.~\ref{#1}}
\newcommand{\eq}[1]{Eq.~(\ref{#1})}
\newcommand{\eqs}[2]{Eqs.~(\ref{#1}) and~(\ref{#2})}

\def\be{\begin{equation}}
\def\ee{\end{equation}}
\def\bea{\begin{eqnarray}}
\def\eea{\end{eqnarray}}
\def\l({\left(}
\def\r){\right)}
\def\Bo{B_{\rm opt}}

\documentclass[twocolumn,ams,preprintnumbers,amsmath,amssymb,superscriptaddress,prb]{revtex4} 
\usepackage{graphicx}
\begin{document}

%------------------------------------------------------------------------------
%\title{Temperature and field distributions
%in superconducting trapped-field magnets during pulsed field activation}

\title{Superconducting trapped-field magnets: Temperature and field 
distributions during pulsed field activation \vspace{-0.0cm}}

\author{S. Br{\ae}ck}
\affiliation{Department of Physics, University of Oslo, PO Box 1048
  Blindern, 0316 Oslo, Norway}
\author{D. V. Shantsev}
\affiliation{Department of Physics, University of Oslo, PO Box 1048
  Blindern, 0316 Oslo, Norway}
\affiliation{A. F. Ioffe  Physico-Technical Institute  of Russian
  Academy of Sciences, 194021
  St. Petersburg, Russia}
 \author{T. H. Johansen}
\affiliation{Department of Physics, University of Oslo, PO Box 1048
  Blindern, 0316 Oslo, Norway}
 \author{Y. M. Galperin}
\affiliation{Department of Physics, University of Oslo, PO Box 1048
  Blindern, 0316 Oslo, Norway}
\affiliation{A. F. Ioffe  Physico-Technical Institute  of Russian
  Academy of Sciences, 194021
  St. Petersburg, Russia}
 
%\date{\today}

~

\vspace{-1.1cm}
\centerline{\hspace{-9cm}Submitted to J. Appl. Phys. on 10.05.2002}

\begin{abstract}

%\vspace{0.3cm}
We calculate the temperature and magnetic field distributions in a 
bulk superconductor during the process of pulsed-field magnetic activation. 
The calculations are based on the heat diffusion equation with 
account of the heat produced by flux motion, and the critical state model 
with temperature dependent critical current density.
For a given activation time the total amount of trapped flux $\Phi$ is  
maximum for an optimal value $B_{\rm opt}$ of the maximal applied field. 
We analyze how $B_{\rm opt}$ and $\Phi$ depend on the material parameters 
and the field ramp rate.
\end{abstract}

\maketitle

%\narrowtext

\section{Introduction \label{sec:intro}}

The recent progress in fabrication of large sized
high-temperature superconductors with high critical current density
makes them extremely promising for use as 
permanent magnets. Trapped fields exceeding 12~T have already been reported
for a YBa$_2$Cu$_3$O$_{7-\delta}$ magnet at 22~K.\cite{gruss}
To magnetically activate the superconductor one most often uses a pulsed field magnetization (PFM).
Whereas the PFM activation method is the most convenient 
from practical point of view, even higher fields 
have been trapped by quasi-static field ramping.\cite{ikuta00,surzhenko,ikuta01,mizutani}
This shortcoming of PFM cannot be overcome by simply increasing 
the maximal applied field $B_M$.
In fact, it is found experimentally that the trapped flux reaches maximum at some optimal value $B_M=\Bo$,
and decreases for larger $B_M$.\cite{itoh,itoh-2,
surzhenko,ikuta00,sander,ikuta01,mizutani}
This behavior is believed to result from heating produced by flux motion,
which leads to strong temperature rise in superconductor during the activation process.      
There have already been suggested practical ways to improve the situation, in particular,
by applying multiple field pulses.\cite{sander,ikuta01,sander02}
Even better results can be expected if the development is accompanied by
a modeling of how the heat actually dissipates and 
redistributes in a superconductor during the PFM. Surprisingly, almost no efforts 
have been done in this direction, and this lack of insight motivates the present work.

Our analysis is based on the heat diffusion equation, 
taking into account the time and position dependent heat
dissipation due to flux motion.   
As a result, the temperature and magnetic field 
distributions in the superconductor during all the stages
of the PFM process are calculated.
We determine the optimal applied field $\Bo$ corresponding to the maximum trapped flux
$\Phi(\Bo)$ and analyze how $\Bo$ and $\Phi(\Bo)$ depend on the parameters
of superconductor.

\section{Equations and basic approach}

The distributions of magnetic field, ${\bf B}$, electric field ${\bf E}$ 
and current density ${\bf j}$ inside a sample are determined by
the Maxwell equations
\begin{eqnarray}
\nabla \times {\bf B} &=& \mu_{0}{\bf j}\, ,   \label{eq:Maxwell-Bj} \\
 \nabla \times {\bf  E} &=& - \partial {\bf B}/\partial t\, .
 \label{eq:Maxwell-EB} 
\end{eqnarray}
These equations should be supplemented by a relationship between ${\bf
  j}$ and the fields ${\bf E}$ and ${\bf B}$, which depends on the
  superconductor material, as well as on temperature $T$.  
When a superconductor is subjected to a non-stationary external
magnetic field, $B_a (t)$, a heat per unit volume is produced with the rate 
\begin{equation}
  W = {\bf E} \cdot {\bf j}\, .   \label{eq:Heat}
\end{equation}
The heat release creates a temperature rise
which is given by the thermal diffusion equation
\begin{equation}
C\, \partial T/\partial t - \kappa \nabla^{2} T = W
\, .\label{eq:Temp}
\end{equation}
Here  $\kappa$ is the thermal conductivity, and $C$ the heat capacity per
volume. 

For simplicity we will make calculations for a superconductor shaped as a slab,
while we expect that all qualitative results are valid also for
other geometries. The superconductor occupies space
$|x| \le w$, and satisfies the boundary condition $T(|x|=w)=T_0$, which assumes
ideal thermal contact with the surroundings at the temperature
$T_0$. The solution for a uniform initial temperature can be expressed as 
\begin{equation}
  T(x,t)=T_0 + \int_{-w}^{w} dx' \int_0^t dt' \, G(x,t;x',t') \,
  \frac{W(x',t')}{C}  , \ \label{greens_function} 
\end{equation}
where 
 $G(x,t;x',t')$ is the Green's function due to a unit 
instantaneous plane source at $x'$ at $t'$. The Green's function
satisfying the boundary conditions $G(x=\pm w,t;t',x')=0$
is given by~\cite{Car-Jae} 
\begin{eqnarray}
G(x,t;x',t') = 
\frac{1}{w}\sum_{n=1}^{\infty} \left(
\cos^2\frac{n\pi}{2}\sin\frac{n\pi x}{2w}\sin\frac{n\pi x'}{2w} 
\right. \nonumber \\
  \left.  + \sin^2\frac{n\pi}{2}\cos\frac{n\pi x}{2w}
\cos\frac{n\pi x'}{2w} \right) \,
e^{-\kappa n^2\pi^2(t-t')/4Cw^2} .  
\label{func:Green}
\end{eqnarray}

To calculate 
heat release 
rate $W(x,t)$ we 
use the critical state model, according to
which the magnitude of the current density in flux-penetrated regions of a
superconductor equals to some critical value, $j_c$. The critical current density
depends in general on both the local field $B$ and local temperature
$T$, thus the magnetic field profile is determined by 
Eq.~(\ref{eq:Maxwell-Bj}) with $|{\bf j}|=j_c(B,T)$. 
In order to obtain analytical results, we choose 
the Bean model, i.~e. assume $j_c$ to be
$B$ independent. To account for the $T$ dependence of
$j_c$ the following iterative procedure is used. First, 
$j_c$ is taken $T$-independent and the time evolution of
the profiles $B(x,t)$ and $W(x,t)$ is calculated. The $W$
is substituted in Eq.~(\ref{greens_function}) to determine the
temperature profile $T(x,t)$. It is then used to recalculate the
$j_c[T(x,t)]$ which subsequently gives corrected
magnetic field profiles according to \eq{eq:Maxwell-Bj}.
Fortunately, it turns out that even the 1st
iteration gives correct results within very good accuracy for realistic
parameters. This is demonstrated below by a self-consistent
numerical solution of the above set of equations.

Let us consider a PFM where the external field is applied as a triangular pulse,
\begin{equation}
B_a(t) = R \ (t_M-|t-t_M|)\, ,  
\label{func:Ap-field}
\end{equation}
where $R$ is a constant ramp rate, i.~e. the field increases 
to the maximum value $B_M$ during $0 \le t \le
t_M=B_M/R$, and then decreases to zero
during $t_M < t \le 2t_M$.
When the temperature dependence of the critical current is neglected the
field profile is given by the conventional Bean model
for a zero-field-cooled superconductor, i.e., 
when the external field \emph{increases}, the magnetic flux
occupies the region $x_0 (t) \le |x| \le w$, where
\begin{equation}
  \label{x-zero}
  x_0(t)=w-vt\, , \quad v=R/\mu_0 j_{c0}\, , \quad  j_{c0} \equiv j_c(T_0)
  \, .
\end{equation}
The heat release in the penetrated region is obtained from
Eqs.~(\ref{eq:Maxwell-EB}) and (\ref{eq:Heat}): 
\begin{equation}
  \label{func:partQ2}
  W(x,t)=\left[|x|-x_0(t)\right]Rj_{c0}
\end{equation}
while $W(x,t)=0$ at  $|x| < x_0 (t)$. After the sample becomes fully
penetrated, i.~e. at $t \ge w/v$, one has
\begin{equation}
W (x,t) =  |x| Rj_{c0} \, . \label{func:fullQ2}
\end{equation} 
Similarly, for {\em decreasing} applied field we obtain
\begin{equation}
W(x,t) = 
              \left[|x|-x_1(t)\right]Rj_{c0} \, , \quad x_1(t)\le |x| \le w  \, ,
         \label{func:partQ2a}
\end{equation}
where 
\begin{equation}
x_1(t)=w-v(t-t_M)/2  
\label{x1}
\end{equation}
is the position of maximum flux density.
In the region $|x| < x_1(t)$ there is no flux motion, and therefore
no heat release. For $t > t_M+2w/v$ the field
is decreasing throughout the sample, and the dissipation is again given by
Eq.~(\ref{func:fullQ2}). {}From this set of $W(x,t)$ one finds
the temperature distributions $T(x,t)$
at all stages of the process. The expressions are listed in the Appendix.
Finally, we determine the refined $B(x,t))$  
by assuming a linear $T$ dependence of the critical current,  
\begin{equation}
j_c(x,t) =j_{c0}\ \frac{T_c-T(x,t)}{T_c-T_0} \label{func:S-curr}
\end{equation}
where $T_c$ is the critical temperature.\cite{endnote}  

The present thermo-magnetic problem is characterized by only two dimensionless parameters. The
first is the ratio of $t_M$ to the thermal diffusion time $Cl^2/\kappa$,
\begin{equation}
  \label{tau}
  \tau=t_M \kappa/Cl^2 = B_M \kappa/RCw^2\, .
\end{equation}
If $\tau \ll 1$ the heat diffusion can be neglected, 
whereas for $\tau \gg 1$ the heat escapes the sample so fast
that the temperature increase is negligible.
The second parameter is 
\begin{equation}
  \label{alpha}
  \alpha=\frac{B_p}{B_{fj}}= \left(\frac{\mu_0 w^2 j_{c0}}{2C}\,
  \frac{\partial j_c}{\partial T} \right)^{1/2}\, . 
\end{equation}
Here $B_p=\mu_0j_{c0}w$ is the full penetration field, and $B_{fj}$ is
the threshold field for a flux jump.~\cite{wipf67,wipf}
Consideration of flux jumps -- macroscopic flux avalanches  accompanied by
pronounced heating -- is beyond the scope of the present study.
In practice, one wants to avoid flux jumps, which ruin the magnetization process, 
and can even damage the material.
Therefore we limit ourselves to the parameter range where $\alpha <1$, and
where the above iteration procedure is applicable.

\section{Results and discussion \label{discussion}}
\begin{figure}
\begin{center}
  \includegraphics[width=8cm]{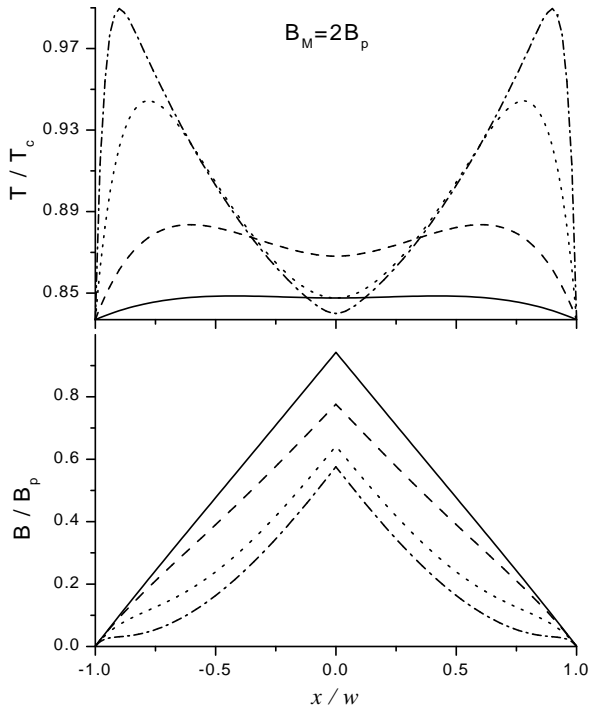}
\end{center}

\vspace{-0.3cm}
\caption{Temperature and flux density profiles in the
remanent state after application of magnetic field $B_M=2B_p$.
Four curves are calculated for different durations of the field pulse: 
%$R=1.43$~T/s (solid), 14.3~T/s (dashed), 143~T/s (dash-dotted) and 1430~T/s (dotted), 
$2t_M=7.3$~s (solid), 0.73~s (dashed), 73~ms (dotted) and 7.3~ms (dash-dotted), 
which correspond to $\tau=1$, 0.1, 0.01 and~0.001, respectively.  }
\label{fig:2-rems}
\end{figure}
\begin{figure}
\begin{center}
  \includegraphics[width=8cm]{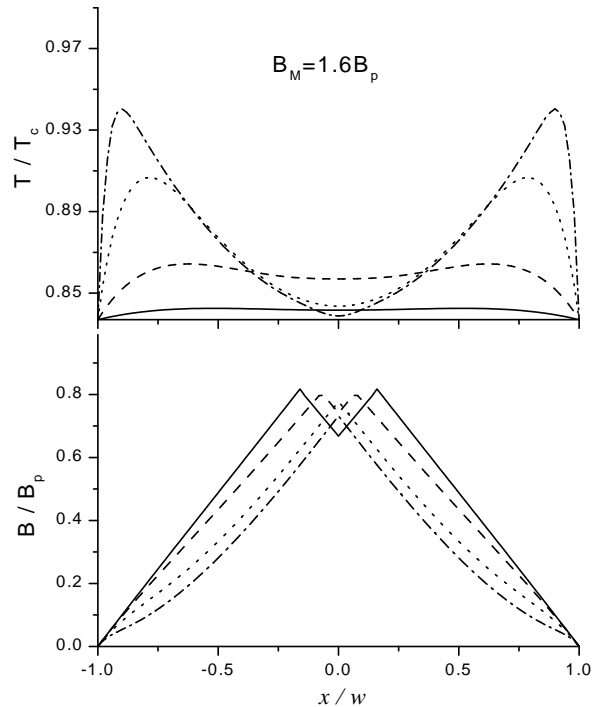}

\vspace{-0.3cm}
\end{center}
\caption{The same curves as in \f{fig:2-rems} but for 
the maximum applied field $B_M=1.6B_p$. }
\label{fig:1.5-rems}
\end{figure}
\begin{figure}
\begin{center}
  \includegraphics[width=8cm]{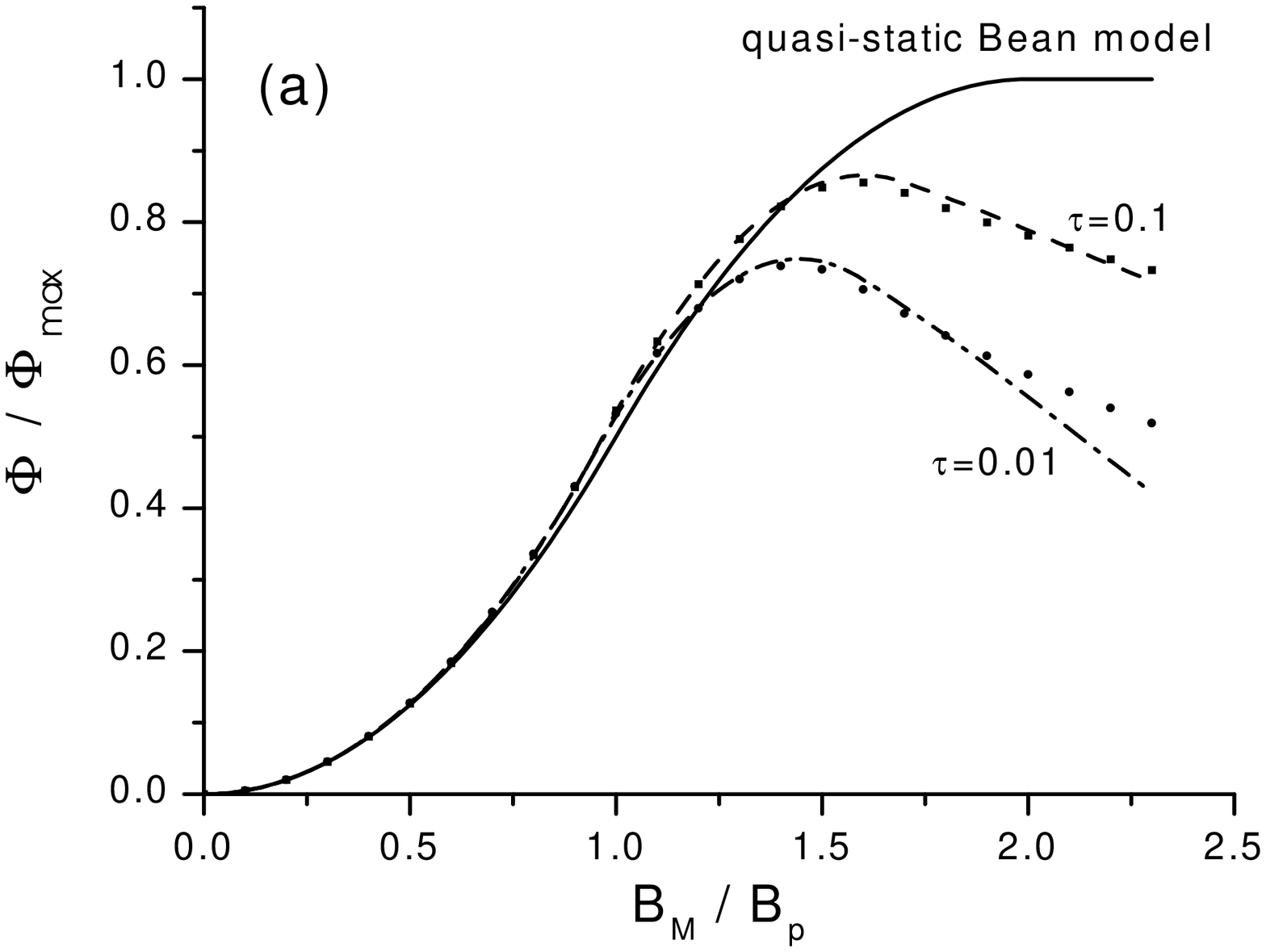}
  \includegraphics[width=8cm]{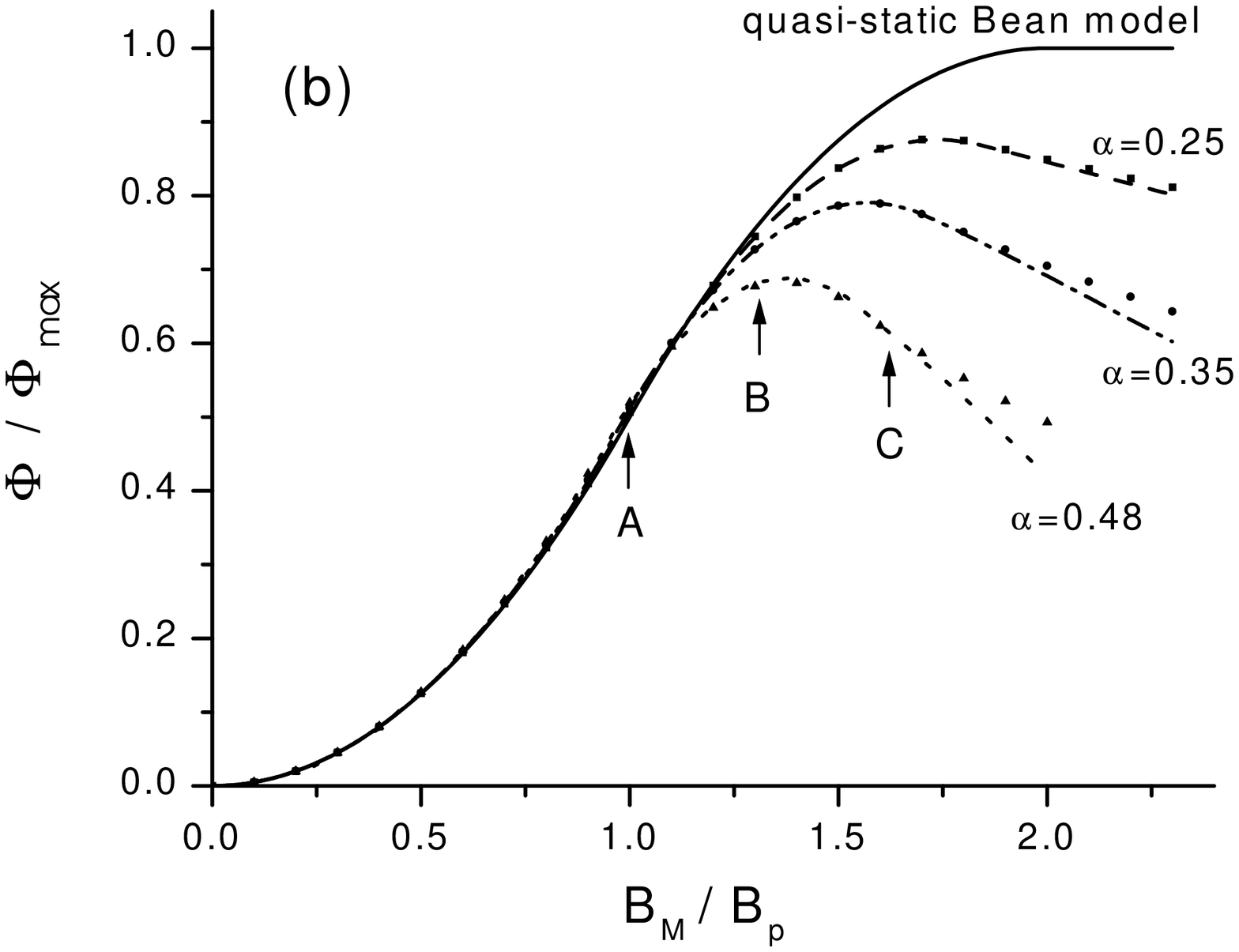}
\end{center}

\vspace{-0.2cm}
\caption{Trapped remanent flux after a magnetization process for
different maximum applied fields $B_M$. 
(a) Curves for $\alpha=0.48$ and different $\tau$, i.~e.,
for different rates.
(b) Curves for $\tau=0.001$ and different $\alpha$, i.~e.,
for different heat capacities.
The solid curve is the same for both plots and shows
the Bean-model result without any heating effects.
Symbols show results of numerical calculations which
reproduce well our analytical curves except for
very large $B_M$. }
\label{fig:remflux}
\end{figure}
\begin{figure}
\begin{center}
  \includegraphics[width=8.4cm]{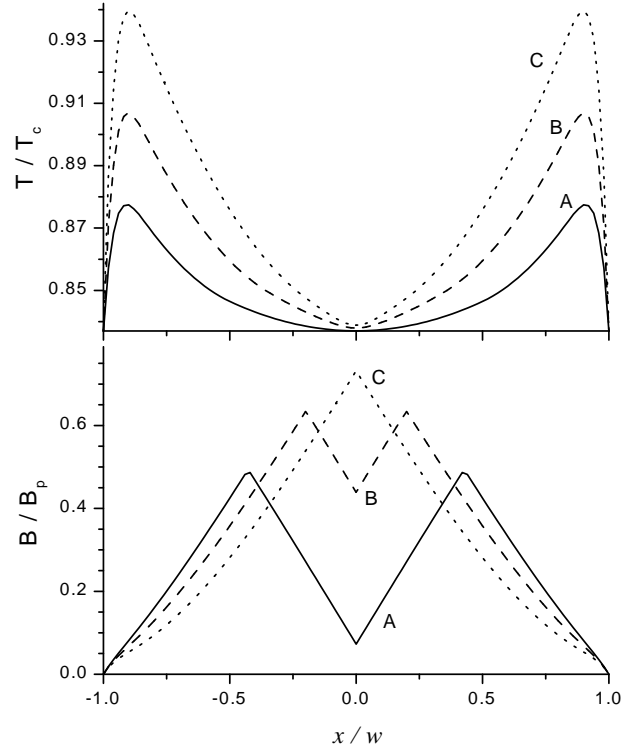}
\end{center}

\vspace{-0.3cm}
\caption{Temperature and flux density profiles in the remanent
state after application of the maximum field $B_M=B_p$ (A), 
$B_M=1.3B_p$ (B), and $B_M=1.6B_p$ (C).  
The profiles correspond to the marked points (A,B,C) on the lowest curve
of \f{fig:remflux}(b). The profile (B) corresponds to the maximum trapped flux.
%The corresponding applied field
%sweeping rates are $715$ T/s, $929$ T/s and $1144$ T/s, respectively.
%The duration of the process is $7.74$ ms.
}
\label{fig:F-rems}
\end{figure}
\begin{figure}
\begin{center}
  \includegraphics[width=7.6cm]{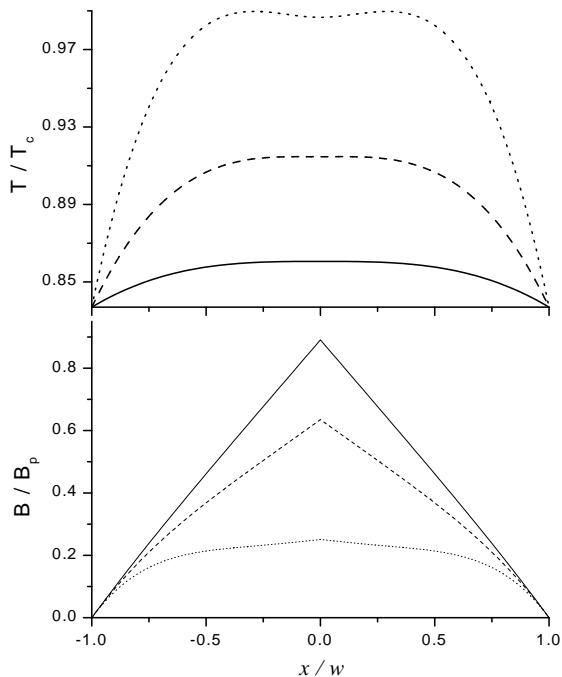}
\end{center}

\vspace{-0.3cm}
\caption{Temperature and flux density profiles in the
remanent state after applying magnetic field 
$B_M=6B_p$.  
%The curves are obtained for relatively slow field sweep:
%$R=1.43$ T/s (solid), 5.3~T/s (dashed), and 14.3~T/s (dotted),
The curves are obtained for the field-pulse duration:
23.2~s (solid), 6.3~s (dashed), and 2.3~s (dotted),
which correspond to $\tau$=3.2, 0.9 and 0.32, respectively.}
\label{fig:6-rems}
\end{figure}
\begin{figure}
\begin{center}
  \includegraphics[width=8cm]{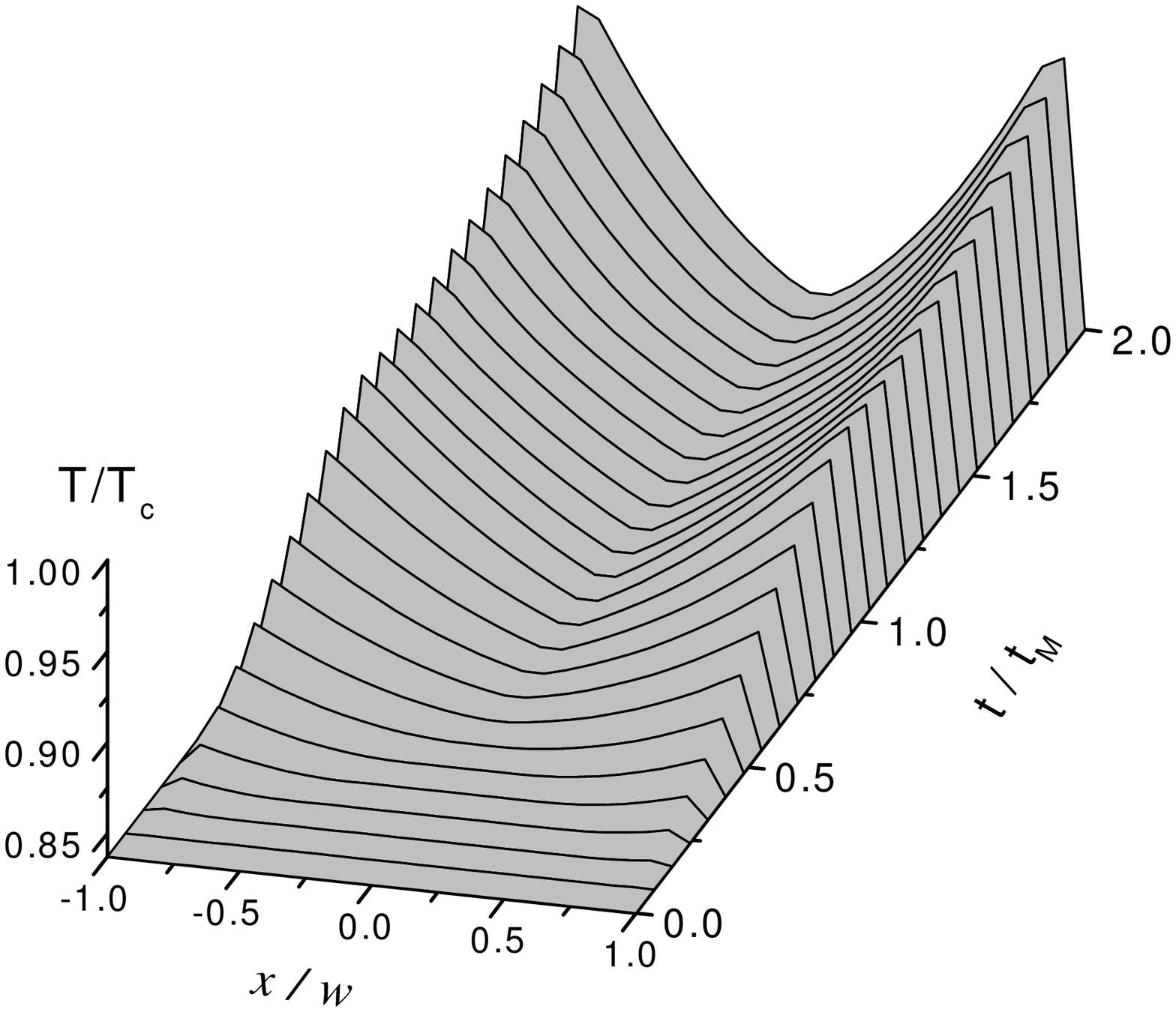}\\
  \includegraphics[width=8cm]{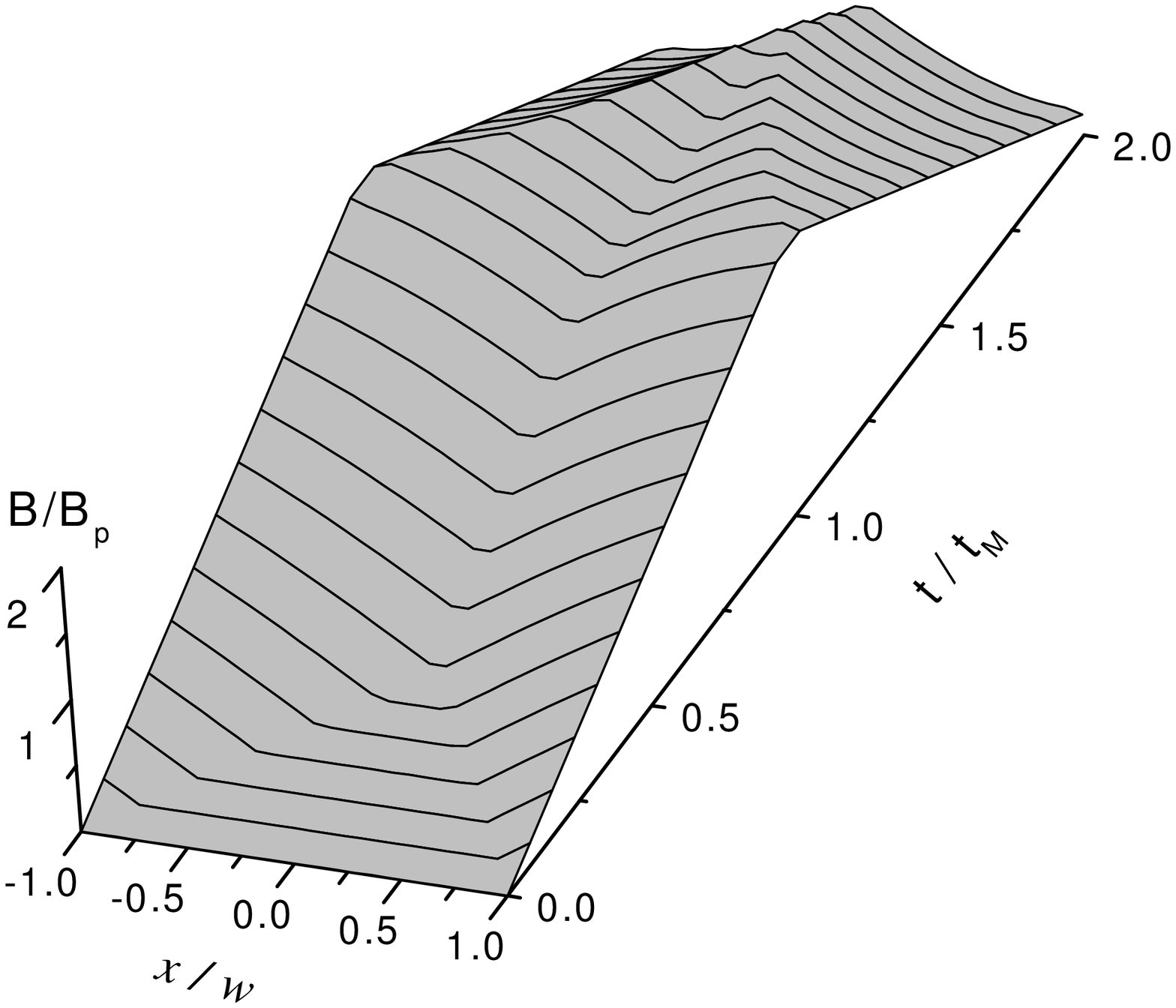}
\end{center}

\vspace{-0.3cm}
\caption{Time evolution of temperature and flux density distrbutions
during the full magnetization
process for the maximal field $B_M=2B_p$, and duration 7.3~ms ($\tau$=0.001).}
\label{fig:3d}
\end{figure}

We present results of the calculations assuming a sample of size $2w=1$ cm
and characterized by  
$j_{c0}=4.4\cdot 10^8$ A/m$^2$ (giving $B_p\approx 2.8$~T),  
$C=0.88 \times 10^6$ J/m$^3$K, and $\kappa =6$ J/msK, 
all of which are typical values for YBa$_2$Cu$_3$O$_{7-\delta}$
crystal with $T_c=92$~K at $T_0=77$~K.
{}From Eq.~(\ref{func:S-curr}) one finds that the above set of
parameters corresponds to $\alpha=0.48<1$. 

Shown in Figs.~\ref{fig:2-rems} and~\ref{fig:1.5-rems} are the
temperature and field distributions after completed PFM, i.e., in the remanent state 
at $t=2t_M$. The figures display the result for the cases $B_M=2B_p$ and $B_M=1.6B_p$, 
and the profiles are plotted for four different ramp rates of the
applied field, corresponding to $\tau$ ranging from 1 to 0.001.
% We consider here a sample with $2l=1$ cm
% characterized by 
% %$\rho =5500$ kg/m$^3$, $C=160$ J/kg$\cdot$K,
% $C=0.88 \times 10^6$ J/m$^3$K ,  
% $j_{c0}^s=4.4\cdot 10^8$ A/m$^2$, % $\kappa =6.82\cdot 10^{-6}$
%                                   % m$^2$/s 
% $\kappa =6$ J/msK, 
% and 
% $T_c=92$ K which are typical values for YBaCuO at $T_0=77$ K. 
%As follows from the upper panels of Figs.~\ref{fig:2-rems} and
%\ref{fig:1.5-rems}, 
At high rates the temperature profiles have pronounced maxima near the
sample surface. The reason is obvious -- near the surface
the flux motion is most intense and the heat release is maximal, see
Eqs.~(\ref{func:partQ2})-(\ref{func:partQ2a}). For a high
ramp rate ($\tau \ll 1$), the heat remains mostly
in the regions where it was released. 
If the rate is low ($\tau \gtrsim 1$) the heat has 
time to diffuse both to the center and the surface. 
As a result, $T(x,t)$ becomes more uniform, but on the average 
the temperature rise is smaller than for high rates.  

As seen from the lower panels 
the temperature rise strongly affects the profiles of the remanent 
field. Increased temperatures give a lower critical current, and
hence, less steep slopes $|\partial B/\partial x|=\mu_0
j_{c}(T)$. For the highest ramp rate
the profiles are also most non-linear, and the total amount of 
trapped flux is lowest. Hence, we reproduce the expected result
that very low ramp rates give maximum trapped flux. 
 
In practice, the duration of the field pulse is limited.
The key point is then to choose the optimum applied field $B_M$
so that in the remanent state the trapped flux becomes maximum. 
If $B_M$ is too small the flux penetration is also
small, whereas too large $B_M$ gives an excessive heating
and little flux becomes trapped by the sample. 
This is illustrated in Fig.~\ref{fig:remflux}, where
the total amount of trapped flux $\Phi=\int B(x)dx$ per unit length
of the slab is plotted as a function of $B_M$.
The solid curves show the conventional Bean-model result,%\cite{beanF}
\begin{eqnarray}
\Phi & = & w B_M^2/(2B_p), \quad  B_M \le B_p, \nonumber \\
& = & w \left[ 2B_M-B_p-B_M^2/(2B_p) \right] , \quad   B_p \le B_M \le 2B_p, \nonumber \\
& = & w B_p, \quad B_M \ge 2B_p,
\end{eqnarray}
obtained for a slab superconductor assuming linear field profiles. 
This result is applicable when the heating
is negligible, i.e., for very low ramp rates or
large $C$ or $\kappa$, i.e., when $\tau \rightarrow \infty$ or
$\alpha \rightarrow 0$. 
The presence of heating leads to reduction of the trapped flux, and 
a peak in the $\Phi(B_M)$ curve appears
at some $B_M$ between $B_p$ and $2B_p$.
The two panels of \f{fig:remflux} allow us
to trace the effect of changing the parameters $\tau$ and  $\alpha$.
Plotted as symbols in \f{fig:remflux} are the exact results obtained 
by numerical simulations, whereas the lines represent the analytical solution.
For the most important range of $B_M$ where $\Phi$ has the peak, the agreement
is excellent, and only a slight deviation appears for large $B_M$.
We conclude therefore that the model given by Eqs.(\ref{greens_function}), 
(\ref{func:Green}), (\ref{func:partQ2})-(\ref{func:partQ2a})
allows to determine the $B_M$ producing the maximum trapped flux.
 
Let us consider 3 characteristic points, A, B and C on the $\Phi(B_M)$ 
where the peak is most pronounced, and analyze the corresponding $B$ and 
$T$ distributions. These remanent profiles are shown in \f{fig:F-rems}. 
For small $B_M$ (A) too little flux penetrated 
the superconductor, and the $B(x)$ profile has a large dip 
in the center region. For large $B_M$ (C), the $B(x)$ profile has the ``right''
triangular shape, however, its slopes are not maximally steep due to heating.
In the state of maximum trapped flux (B) the small dip in the center
is compensated by having relatively large slopes in the overall peak.
Interestingly, we find that the optimum case always has a small minimum in 
the flux density at the center of the superconductor.

If the maximum applied field is very large, the heat is most of the time 
released uniformly throughout the superconductor. Meanwhile, 
the heat is removed only through the surface which is maintained at a 
fixed temperature. 
As a result, the remanent $T(x,2t_M)$ has a broad maximum in the center, and 
the trapped $B(x,2t_M)$ acquire a specific ``bell'' shape. 
This is illustrated in \f{fig:6-rems} for the case $B_M=6B_p$. 

It is also interesting to analyze the evolution of the temperature
and flux density during the whole magnetization process.
A time sequence of curves showing this is plotted in \f{fig:3d}, and corresponds
to the intermediate stages leading to the remanent state seen in 
Fig.~\ref{fig:2-rems} for $\tau = 0.001$.
One can see that the temperature growth starts slowly and
then accelerates. When the applied field reaches maximum 
(at $t=t_M$), the $T$ profile has already acquired
its characteristic shape and changes little during the subsequent field decrease.
The evolution of the $B$ profile looks similar to the standard
Bean-model behavior. The full penetration is reached
approximately at $B_a=B_p$ (at $t=t_M/2$), and then 
the $B(x,t)$ shifts upward almost uniformly until the field starts to decrease
and flux exits from the surface.
The flux front position (seen as a ridge
in the 3D plot) is shifting towards the center, and in the final 
remanent state the $B$ profile acquires its triangular shape.   

Our approach, used so far for a constant ramp rate of the field,
is easily generalized to having different sweep rates
on the ascending and descending field branch.
It turns out that for a given magnetization time it is somewhat beneficial
to have a faster field increase followed by a slower decrease to zero.
In particular, for the extreme case of an instantaneous
field increase and a descent lasting 8~ms we find that one traps 8\% more flux
than for the symmetric 4+4~ms field pulse (for the parameters used in the paper).
The physical reason for this is the following.
When a larger amount of heat is dissipated
in the beginning of the PFM, the heat has more time 
to flow out of the sample.
As a result, the temperature in the remanent state
becomes slightly lower, which results in a larger trapped flux.  
We omit to show detailed profiles, since these graphs are very similar to the ones 
already presented in this paper.
  
The critical state model used in our analysis 
neglects the viscous force acting on flux lines. 
This force can be an important ingredient that
determines the $B(x,t)$ distributions during very fast PFM.\cite{itoh}
We stress, however, that the heating takes place independent of
whether the flux motion is viscous or not. 
Our results clearly demonstrate that the heating produced 
within the critical-state approach is sufficient to account for
the suppression of the trapped flux during the PFM process.
   
\section{Conclusion \label{conclusion}}

The temperature and field distributions in a bulk superconductor 
during a PFM process are calculated
analytically within the critical state model and 
with account of heating due to flux motion.
The remanent trapped flux $\Phi$ is smaller for a larger PFM rate, 
and for smaller heat capacity.
For a given duration of the activation pulse the $\Phi$ reaches maximum 
for some optimal maximum field, which is always smaller than twice
the penetration field. 
Suprizingly, the remanent flux distribution for 
optimal field is not monotonous,
but the overall peak has a small dip in the center.
The strongest temperature rise is usually found close to the
surface. The trapped flux 
can be enhanced without changing the total PFM time if 
the field ascent is made faster than the descent to zero.

\begin{table}
\centerline{
\begin{tabular}{|c|rcl|c|} \hline
 %$B$ & Time interval & \#\\ \hline
$B_M \leq B_p$ &  & $t$ & $\leq t_M $ & (\ref{1a})\\
               & $t_M <$ &$t$&$ \le t_M+B_M/R_D$ & (\ref{1b})\\ \hline
               &  &$t$&$ \leq B_p/R_A $ & (\ref{1a}) \\
$B_p \le B_M \le 2B_p$   &$B_p/R_A <$ &$t$&$\leq t_M$& (\ref{2b})\\
               &$t_M <$ &$t$&$ \le t_M+B_M/R_D$& (\ref{3c})\\ \hline
               & &$t$&$ \leq B_p/R_A $ & (\ref{1a}) \\
 $B_M \ge 2B_p$   & $B_p/R_A <$ &$t$&$ \leq t_M$ & (\ref{2b}) \\
               &$t_M <$ &$t$&$ \leq t_M+ 2 B_p/R_D$ & (\ref{3c}) \\
              & ~$t_M+ 2 B_p/R_D <$ &$t$&$ \le t_M +B_M/R_D$& (\ref{a5})\\ \hline
\end{tabular} 
}\caption{The last column shows which formula for coefficients $S_n$ in \eq{t}
should be used for given $B_M$ and $t$.} 
\end{table}

\appendix
\section{}

Here we list analytical expressions for the temperature distribution 
for different
stages of the magnetization process. 
They are obtained by substituting the heat release rate given by
Eqs.(\ref{func:partQ2})-(\ref{func:partQ2a})
and the Green function (\ref{func:Green}) into 
\eq{greens_function}.
The expressions are given for a general case of having different field
ramp rates $R_A$ and $R_D$ 
on the ascending and descending branch, respectively.
The following notations are used:
$a= \pi^2 \kappa/4Cw^2$ (inverse thermal diffusion time), 
$b=%2Cl^2/\pi \kappa B_p
\pi/(2aB_p)$, and $t_M$ is the time when the applied field  
reaches maximum, $t_M=B_M/R_A$. 
The temperature distribution during the magnetization process is given by
\begin{equation}
T(x,t) = T_0 + \frac{32j_{c0}w^3 R_A}{\pi^4\kappa} \sum_{n=1}^{\infty}
S_n(t) \; \sin\frac{n\pi}{2}\cos\frac{n\pi x}{2w}  
\label{t}
\end{equation}
The dimensionless coefficient $S_n$ are given by one of the expressions below,
and Tab.~1 explains which expression should be used for every 
particular case. It is convenient to define the following functions
corresponding to incomplete and complete penetration processes:
\begin{eqnarray}
&&S_n^{\rm inc}(t,R)  \equiv 
 \frac{(an^2t+e^{-an^2t}-1) bR}{n^5}  + \nonumber \\  
&& \frac{\left[\cos (abRnt) -e^{-an^2t}\right]bR/n
 -\sin (abRnt)}{n^2\left[n^2+\left(bR\right)^2\right]} \  ,  
 \label{inc} 
\end{eqnarray}
\begin{widetext}
\begin{equation}
S_n^{\rm com}(t,R)= 
\frac{e^{-an^2(t-B_p/R)}}{n^3}
\Bigg\{ \frac{\pi}{2}-\frac{bR}{n^2}+
\frac{(bR)^3 \ e^{-an^2B_p/R} - n^3 \sin\frac{n\pi}{2}} 
{n^2 \left[n^2+(bR)^2\right]}
+ \left(\frac{\pi}{2} \sin\frac{n\pi}{2} - \frac{1}{n}\right)
\sin\frac{n\pi}{2}
\left(e^{an^2(t-B_p/R)}-1\right)
 \Bigg\}\, ,
\label{sncom}
\end{equation}
Then, 
\begin{eqnarray}
% ********** 1
S_n&=&S_n^{\rm inc}(t,R_A) \, , 
 \label{1a} \\ 
% ********** 2
S_n&=&  
 e^{-an^2(t-t_M)} S_n^{\rm inc}(t_M,R_A)
 + \frac{R_D}{R_A}\ S_n^{\rm inc}(t-t_M,R_D/2)
 \, ,  \label{1b}\\
% *********** 3
S_n &= &S_n^{\rm com}(t,R_A) \, ,
\label{2b}\\
% ************ 4
S_n&=& e^{-an^2(t-t_M)} S_n^{\rm com}(t_M,R_A)   + \frac{R_D}{R_A}\ S_n^{\rm inc}(t-t_M,R_D/2)
 \, . \label{3c} \\
% *************** 5
S_n&= &e^{-an^2(t-t_M)}  S_n^{\rm com}(t_M,R_A)  +  \frac{R_D}{R_A}\ S_n^{\rm com}(t-t_M,R_D/2) 
\label{a5}
\end{eqnarray}
\end{widetext}

%\widetext

\begin{references}

\bibitem[*]{0} Email: t.h.johansen@fys.uio.no

% 16T
\bibitem{gruss} S. Gruss, G. Fuchs, G. Krabbes, P. Verges, 
G. St\"{o}ver, K.-H. M\"{u}ller, J. Fink,
and L. Schultz, Appl. Phys. Lett. {\bf 79} (2001) 3131.

\bibitem{mizutani} U. Mizutani, H. Ikuta, T. Hosokawa, H. Ishihara, K. Tazoe,
T. Oka, Y. Itoh, Y. Yanagi and M. Yoshikawa, 
Supercond. Sci. Technol. {\bf 13} (2000) 836.

% B(x,t)
\bibitem{ikuta00} H. Ikuta, H. Ishihara, T. Hosokawa, Y. Yanagi,
Y. Itoh, M. Yoshikawa, T. Oka and U. Mizutani,  
Supercond. Sci. Technol. {\bf 13} (2000) 846.

\bibitem{ikuta01}  H. Ikuta, Y. Yanagi, H. Ishihara, M. Yoshikawa,  
Y. Itoh, T. Oka and U. Mizutani, Physica C {\bf 357} (2001) 837.

\bibitem{surzhenko} A. B. Surzhenko, S. Schauroth, D. Litzkendorf, M. Zeisberger,
T. Habisreuther and W. Gawalek, Supercond. Sci. Technol. {\bf 14} (2001) 770.

%multipulse:
\bibitem{sander} M. Sander, U. Sutter, R. Koch and M. Kl\"{a}ser,
Supercond. Sci. Technol. {\bf 13} (2000) 841. 

% main two
\bibitem{itoh} Y. Itoh and U. Mizutani, Jpn. J. Appl. Phys. {\bf 35} (1996) 2114.
\bibitem{itoh-2} Y. Itoh, Y. Yanagi, M. Yoshikawa, T. Oka, 
Y. Yamada, and U. Mizutani, Jpn. J. Appl. Phys. {\bf 35} (1996) L1173.

\bibitem{sander02} M. Sander, U. Sutter, M. Adam and M. Klaser,
Supercond. Sci. Technol. {\bf 15} (2002) 748. 

\bibitem{Car-Jae} H. S. Carslaw and J. C. Jaeger, {\em Conduction of Heat in Solids} 
(Clarendon Press, Oxford, 1959).

\bibitem{endnote} The determination of $B(x,t)$ is straighforward when the magnetic
field either increases or decreases throughout the sample. Otherwise,
there is a kink in the field profile. In the zeroth approximation, the
kink position is given by \eqs{x-zero}{x1} for $x_0(t)$ or
$x_1(t)$, for increasing/decreasing field, respectively. With account
of the position-dependent temperature, these quantities can be obtained
from the equations
\begin{eqnarray}
  \label{x12}
 \mu_0 \int_{x_0(t)}^w j_c[T(x,t)]\,  dx &=&B_a(t) \, ,\nonumber\\ 
% \left(\int_0^{x_1(t)}-\int_{x_1(t)}^l\right) j_c[T(x,t)]\, dx
% &=&B_a(t) -B(0,t_M)\, .
 \mu_0   \int_{x_1(t)}^w j_c[T(x,t)]\, dx&=&B_a(t) -B[x_1(t),t_M]\, . \nonumber
\end{eqnarray}

\bibitem{wipf67} S. L. Wipf, Phys. Rev. {\bf 161}, (1967) 404.
\bibitem{wipf} S. L. Wipf, Cryogenics {\bf 31}, (1991) 936.

\end{references}
\end{document}